# Origin of Suppressed Ferroelectricity in κ-Ga$_2$O$_3$: Interplay Between Polarization and Lattice Domain Walls

Yonghao Zhu[1], Zhi Wang[1*], Junwei Luo[1**], Lin-Wang Wang[1†]

[1]State Key Laboratory of Superlattices and Microstructures, Institute of Semiconductors, Chinese Academy of Sciences, Beijing, 100083, P. R. China

**ABSTRACT**

The large discrepancy between experimental and theoretical remanent polarization and coercive field limits the applications of wide-band-gap ferroelectric materials. Here, using a machine-learning potential trained on *ab-initio* molecular dynamics data, we identify a new mechanism of the interplay between polarization domain wall (PDW) and lattice domain wall (LDW) in ferroelectric κ-phase gallium oxide (Ga$_2$O$_3$), which reconciles predictions with experimental observations. Our results reveal that the reversal of out-of-plane polarization is achieved through in-plane sliding and shear of the Ga–O sublayers. This pathway creates strong anisotropy in PDW propagation, and crucially leads to topologically forbidden PDW propagation across the 120° LDWs observed in synthesized samples. The resulting stable network of residual domain walls bypasses slow nucleation and suppresses the observable polarization and coercive field. These insights highlight the potential for tailoring the ferroelectric response in κ-Ga$_2$O$_3$ from lattice-domain engineering.

∗ Corresponding author, E-mail: wangzhi@semi.ac.cn
∗∗ Corresponding author, E-mail: jwluo@semi.ac.cn
† Corresponding author, E-mail: lwwang@semi.ac.cn

## Introduction

Wide-band-gap ferroelectric (FE) materials exhibit switchable, non-volatile polarization together with high breakdown fields and low leakage currents. These attributes suit high-power devices from energy harvesting[1], sensors[2], spintronics[3], to memory devices[4]. Recent advances have enabled epitaxial growth of wide-gap ferroelectrics on wide-gap semiconductors, e.g., ferroelectric κ-$Ga_2O_3$ on GaN and β-$Ga_2O_3$[5, 6], providing a natural route to monolithic FE-power-device integration. However, these FE materials often suffer from severe discrepancies between theoretical predictions and experimental observations. For instance in κ-$Ga_2O_3$, (1) remanent polarization predicted by first-principles calculations (~ 23 μC/$cm^2$, similar to $ZrHfO_2$)[7, 8, 9] vastly exceeds experimental values (< 8.6 μC/$cm^2$)[10, 11]; (2) coercive fields fitted by *ab-initio* results and Landau-Ginzburg (L-G) theory (~3 MV/cm)[7, 12] are also an order of magnitude greater than measured (< 0.5 MV/cm)[10]. In fact, the conventional Ising-like polarization switching model, assuming rigid out-of-plane displacements, fails to capture κ-$Ga_2O_3$ intrinsic dynamics[7]. These striking discrepancies indicate critical unresolved physical mechanisms, and point to an urgent need to reconcile theory with experiment to optimize FE materials for further applications.

In this work, we bridge this gap using the deep-learning (DL) model trained on data from density-functional theory (DFT) and molecular dynamics (MD). We reveal a nontrivial mechanism in κ-$Ga_2O_3$ dominated by interplay between the polarization domain wall (PDW) and the lattice domain walls (LDWs) observed in synthesized samples. The reversal of out-of-plane polarization is achieved through in-plane sliding and shear of the Ga–O sublayers (Fig.1a-e). We demonstrate that such sliding-driven mechanism leads to strong anisotropy of PDW propagation. More crucially, the propagation of PDW can be topologically blocked by the 120° LDWs, resulting in stable coexistence of residual domain walls. This interplay between PDW and LDW fundamentally alters the polarization switching dynamics by circumventing the slow nucleation processes. Instead, it enables rapid PDW propagation with lower coercive field, albeit at the expense of reduced remanent polarization. Our findings explain the suppressed polarization and coercivity observed experimentally, and provide the strategy to engineer ferroelectric responses by controlling lattice domain sizes in this family of wide-band-gap FE materials.

## Results

### Atomic structure and sliding-driven polarization switching in the primitive cell.

The primitive orthorhombic cell of κ-Ga$_2$O$_3$ (space group $Pna2_1$) contains 16 Ga and 24 O atoms. Our DFT lattice constants (*a*=5.04 Å, *b*=8.63 Å, and *c*=9.24 Å) agree well with the experiments[7, 13]. Notably, the ratio $a/b \approx 1/\sqrt{3}$ suggests compatibility with hexagonal substrates, naturally enabling the formation of intrinsic 120° lattice domains[7]. The crystal consists of four inequivalent Ga–O polyhedral, *i.e.*, one tetrahedral (GaO$_4$) and three octahedral (GaO$_6$), each appearing four times per unit cell (Fig. 1a). Spontaneous polarization arises primarily from the displacement of Ga atoms away from the polyhedral centers. Among these, Ga–O tetrahedra dominate the polarization due to the larger separation of positive and negative charge centers.

Along the *c*-axis, the crystal structure naturally separates into four layers, labeled sequentially as layers 1, α, 2, and β (Fig. 1a,c-e). Oxygen atoms are shared between neighbor layers. Layers 1 and 2 each contain equal numbers of Ga–O tetrahedra and octahedra, whereas layers α and β consist of only Ga–O octahedra. Layers 1 and 2, as well as layers α and β, are symmetrically related by a twofold (C$_2$) rotation symmetry (Supplementary Fig. 1). This symmetry enforces cancellation of any in-plane (001) polarization components among Ga–O polyhedra, resulting in a net spontaneous polarization along the *c*-axis ([001] direction). Details about the atomic coordinates and symmetry conditions are available in Supplemental Note 1.

To reveal the mechanism of FE phase transition, we first investigate the polarization switching in this primitive cell model. Using the nudged-elastic-band (NEB) method[14], we interpolated nine intermediate images between the initial (polarization-up, $+P$) and final (polarization-down, $-P$) structures (Fig. 1b). The $+P$ and $-P$ states are related by reflection symmetry with respect to the (001) plane. Increasing the number of intermediate images does not alter our conclusions.

Remarkably, the lowest-energy pathway for polarization switching proceeds via a nontrivial sliding-like mechanism (Fig. 1c-e) rather than the conventional vertical displacement model. Throughout the switching process, layers α and β remain structurally rigid and undergo primarily translational sliding along [100]. Meanwhile, because layers 1 and 2 share oxygen atoms with layers α and β, they exhibit significant shearing deformation along [100]. The Ga atoms in layers 1 and 2 initially at the tetrahedral site (GaO$_4$, labeled "1" in Fig. 1) move along [100] and ultimately transform into octahedral coordination (GaO$_6$). Conversely, the Ga atoms labeled "2" undergoes the opposite transformation from octahedral to tetrahedral coordination. Intuitively, the orientation of the GaO$_4$ tetrahedron at the initial and final states reverses along the polar c-axis, indicating polarization reversal.

The intermediate paraelectric (PE) phase has $Pbcn$ symmetry (Fig. 1d). The calculated energy barrier for this sliding-like pathway is 0.10 eV per formula unit (eV/f.u.), with a predicted spontaneous polarization of 24.58 µC/cm$^2$ from Berry phase method[7, 8, 9, 15] (Fig. 1b). These theoretical values, however, still significantly exceed the experimentally measured polarization and coercive field[10, 11], due to the limitation of the primitive cell model. Nevertheless, this sliding-driven mechanism represents the intrinsic pathway of polarization reversal, and is consistent with results previously predicted in κ-Ga$_2$O$_3$[7], and also in structurally analogous materials, such as Al$_x$Fe$_{2-x}$O$_3$[10] and ε-Fe$_2$O$_3$[16, 17] which shares the same space group as κ-Ga$_2$O$_3$. The sliding-like polarization reversal is further validated by ab initio molecular dynamics (AIMD) simulations, as detailed in Supplementary Note 2.

To rigorously describe the sliding-driven FE transition, we define the order parameter based on the relative sliding displacement vector $\boldsymbol{u}$ of adjacent layers. Specifically, taking layer 1 atoms as a reference, layer α exhibits opposite sliding displacement vectors along [100] in the $+P$ and $-P$ states (Fig. 1f,g), thus defining two distinct order parameters $+u$ and $-u$. The sign convention for the order parameter $\boldsymbol{u}$ is chosen consistently with polarization $\boldsymbol{P}$ to reflect their coupling. The difference in order parameter between the two polarized states is $(+u) - (-u) = a/3$, where $a$ is the [100] lattice constant.

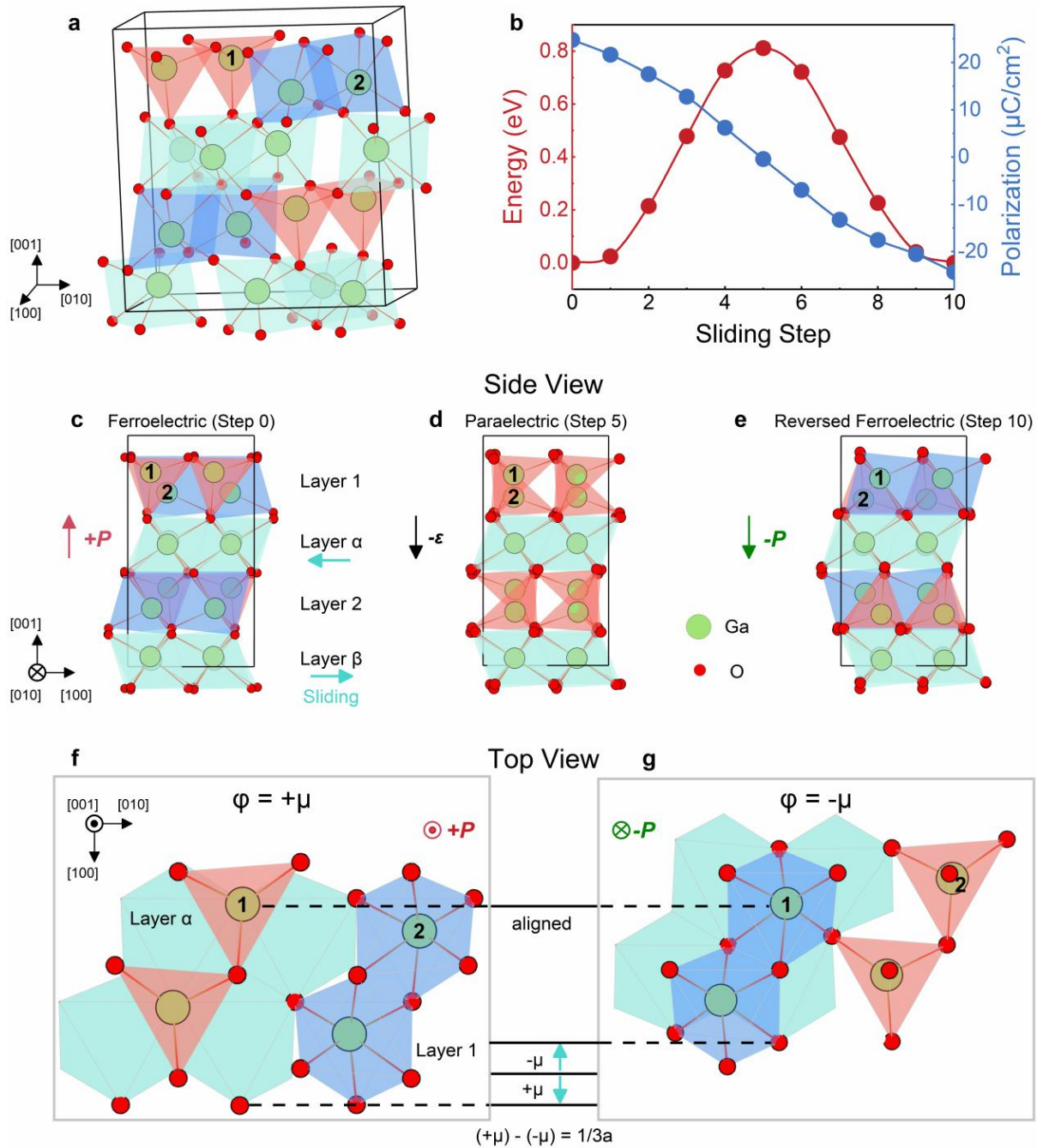

**Figure 1. Structure, and polarization reversal mechanism in κ-Ga₂O₃ primitive cell.** (a) Atomic configurations of the primitive orthorhombic cell, illustrating the four layers along the polar [001] axis and the four inequivalent Ga–O polyhedra (green: Ga, red: O). Spontaneous polarization is indicated by the arrow along [001]. (b) Calculated polarization (via Berry phase method) and energy barrier profiles along the polarization switching pathway obtained by NEB method. (c)-(e) Side views of atomic displacement during the phase transition: (c) initial $+P$ state, (d) intermediate paraelectric state, (d)

final $-P$ state. (f, g) Top views of the $+P$ and $-P$ states, showing the relative sliding vector $\bm{u}$ of layer α along [100]. 3D atomic structures visualized with VESTA software[18].

**The deep learning model.**

The primitive cell model, although a simple and good starting point, lacks the large spatial and temporal scales necessary for critical phenomena in FE phase transition, including the nucleation and growth of polarization domains under finite electric fields[19, 20]. These features often require simulations that exceeds the capability of conventional *ab initio* MD calculations[21]. To overcome this limitation, we apply a deep-learning-based interatomic potential (Deep Potential Long-Range, DPLR) model[22], capturing the long-range electrostatic interactions essential in polar materials[23, 24, 25, 26, 27]. Our model is trained using a dataset comprising 21700 *ab initio* MD configurations, covering a wide range of temperatures (100–1500 K) and ferroelectric switching trajectories driven by external electric fields (see Methods for details).

We validate our DPLR model against DFT calculations, demonstrating excellent predictive accuracy for total energies, atomic forces, and electronic polarization (Fig. 2a–c). Specifically, the root mean square errors (RMSE) are as low as 0.19 meV/atom for energies, 36.61 meV/Å for atomic forces, and 0.0015 Å for WC positions, sufficient to capture the variation of the potential energy surface during the ferroelectric phase transition[28, 29, 30, 31]. The model also reproduces the phonon dispersion in κ-$Ga_2O_3$ ferroelectric phase (Fig. 2d), and predicts Born effective charges (BECs) consistent with direct DFT calculations (Supplementary Fig. 5), both for static configurations and along switching trajectories (Supplementary Note 3).

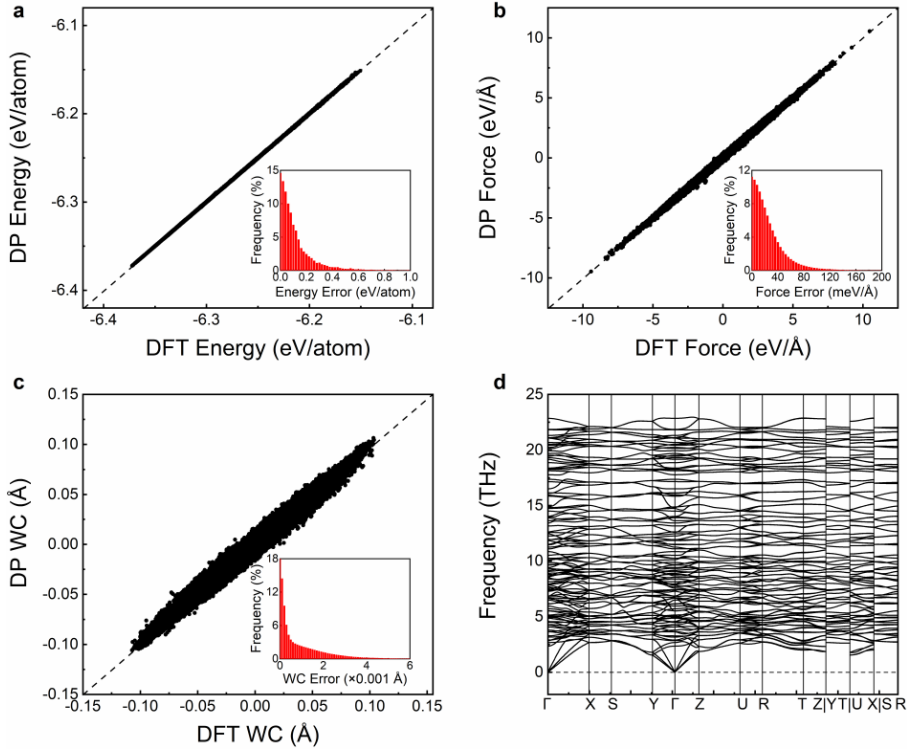

**Figure 2. Accuracy of the DPLR model.** (a-c) Benchmark comparison between DP and DFT for energies, atomic forces, and WC positions, respectively. These WC positions refer to the displacement of the electron centers relative to the corresponding ions. The insert panels are mean absolute error. (d) Phonon dispersion simulated by DP.

The DPLR model is constructed using DeePMD-Kit[32, 33]. The total energy is defined as the sum of short-range energy ($E_{sr}$) and long-range Coulomb energy ($E_{lr}$)[22, 34]. Short-range energy is characterized by a sum of atomic energy, determined by their immediate surroundings with a cutoff radius[35]. Long-range electrostatic interactions between ions (nuclei and core electrons) and valence electrons are approximated via interactions between spherical Gaussian charge distributions centered at ionic positions and maximally localized Wannier centers (WCs)[22, 36]. Moreover, the WCs also is employed to calculate the electronic contribution to polarization ($P$), where the ionic contribution is determined by the ionic charge and positions. Therefore, the total energy under an electric field is given by[37, 38]:

$$E = E_{sr} + E_{lr} - \boldsymbol{P} \cdot \boldsymbol{\varepsilon}_{ext}, \quad (1)$$

where $\varepsilon_{ext}$ is the external electric field. The atomic forces can be calculated by:

$$F = -\frac{\partial E_{sr}}{\partial R} - \frac{\partial E_{lr}}{\partial R} + \frac{\partial P}{\partial R} \cdot \varepsilon_{ext}, \qquad (2)$$

Where $R$ is the atomic position, and $\frac{\partial P}{\partial R}$ is the Born Effective Charge (BEC) tensor[37, 38, 39]. Training data are generated using ab initio MD simulations over temperatures from 100 K to 1500 K, including thermal motions under no external field and trajectories under different external fields[37], resulting in a database of 21700 configurations in total. The machine learning model can predict energy and forces in real time during MD simulations. These configurations can fully describe the potential energy surface of ferroelectric switching. The detailed information can be found in Methods.

## Nucleation and polarization domain wall propagation in single crystal κ-Ga$_2$O$_3$.

We then employ the trained DPLR model to study polarization switching dynamics in a large-scale κ-Ga$_2$O$_3$ single-crystal system containing 9600 atoms (240 unit cells, ~ 10 nm × 10 nm). Initially, the entire crystal is uniformly polarized in the $+P$ state (red region in Fig. 3a). As it is a single crystal, the system contains no lattice domains or lattice domain walls (LDWs), although polarization domain walls (PDWs) can emerge during the dynamics. Under a relatively large external electric field of 24 MV/cm, applied here to accelerate the typically slow nucleation stage (results at experimentally relevant fields are presented later), our simulations clearly capture both the nucleation and PDW propagation stages within a time scale of 3 picoseconds (Fig. 3a–f). Specifically, random thermal fluctuations coupled with the external field initially induce small regions to transform into the metastable PE state (blue regions in Fig. 3b). These metastable regions subsequently nucleate into reversed $-P$ state (green regions in Fig. 3c) upon surpassing a critical nucleus size. The reversed domains then expand and merge through PDW propagation (Fig. 3d,e), eventually leading to complete polarization reversal of the crystal (Fig. 3f).

A notable feature revealed by our simulations is the significant anisotropy in the propagation velocities of PDW. We identify two distinct types of PDWs classified by their orientations: the (100)-oriented PDW and the (010)-oriented PDW, hereafter denoted as PDW$_{(100)}$ and PDW$_{(010)}$, respectively (see Fig. 3d). Quantitative analysis shows that the propagation velocity of PDW$_{(100)}$ (~6975 m/s) is approximately twice that of the PDW$_{(010)}$ (~3451 m/s). This anisotropic behavior can be explained through the variation of the order parameter across the PDWs. For PDW$_{(100)}$, the order parameter changes continuously within the domain wall from $+u$ to $-u$ (Fig. 3g,h). In principle, such continuous variation of order parameter corresponds to a relatively low energy barrier, facilitating rapid PDW propagation. In contrast, PDW$_{(010)}$ involves an abrupt jump in the order parameter due to the structural discontinuity of lattice: The layer α (as well as layer β) is not continuous in the [010] direction, but has periodic "void" regions as shown in Fig. 3i,j. Within PDW$_{(010)}$, the order parameter jumps between $+u$ and $-u$ across these voids. This discontinuity imposes a higher energy barrier, substantially reducing the PDW propagation speed.

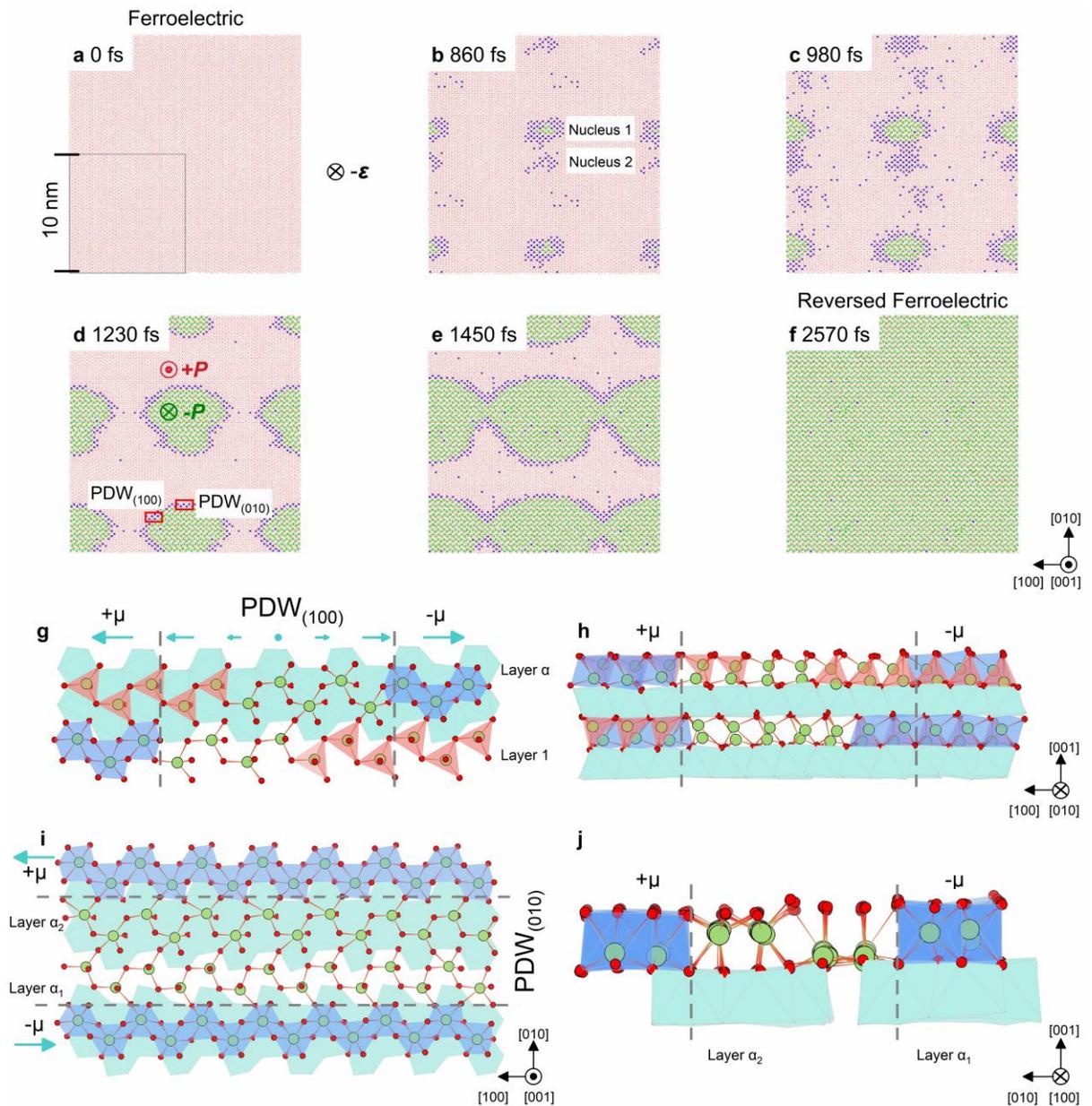

**Figure 3**. Polarization switching dynamics in a 10 nm × 10 nm single crystal of κ-Ga₂O₃. Snapshots (a–f) show the time evolution of the ferroelectric phase transition under an electric field applied along the $[00\bar{1}]$ direction, taken at 0, 860, 980, 1230, 1450, and 2570 fs. Red, blue, and green regions represent the initial +P ferroelectric phase, intermediate paraelectric phase, and reversed –P ferroelectric phase, respectively. In (d), the two representative PDWs - PDW$_{(100)}$ and PDW$_{(010)}$ - are highlighted by red rectangles. Subplots (g-h) and (i-j) show atomic configurations of PDW$_{(100)}$ and PDW$_{(010)}$, respectively. The areas between the two black lines are the regions of domain wall. Order parameters in different areas have been shown by cyan arrows.

Moreover, we observe a clear distinction between the electric field strengths required for nucleation and for PDW propagation. This observation directly connects to two conventional models of ferroelectric switching: the nucleation-limited switching (NLS) model[40], which attributes the switching kinetics to the rate of nucleation, and the Kolmogorov–Avrami–Ishibashi (KAI) model[41], where switching is governed by the propagation of existing PDW[19, 42]. In our simulations, nucleation occurs within the simulation time window (~3 ps) only under a high electric field exceeding 24 MV/cm—more than an order of magnitude higher than typical experimental coercive fields (< 0.5 MV/cm). This helps explain why previous first-principles calculations, which are typically based on small cells and lacked explicit domains, systematically overestimated the coercive field in κ-Ga₂O₃.

Notably, when we initiate simulations from a pre-nucleated configuration containing a reversed domain, a complete polarization switching can occur under much smaller fields as low as 0.2 MV/cm (Fig. 4a), closely matching experimental values. In this case, the polarization reversal is limited by the propagation of PDW$_{(010)}$ which has a slower velocity and higher energy barrier than PDW$_{(100)}$. Fig. 4b shows the relationship between electric field and domain-wall velocity. At high fields ($\varepsilon_{ext} \geq 4$ MV/cm), the extracted velocity-field relation is nearly liner (insert in Fig. 4b), corresponding to the flow region[43]. Under low-field region, the propagation velocity can be described with a creep process[19, 43, 44]:

$$v(\varepsilon_{ext}) = v_0 \exp[-(\frac{\varepsilon_a}{\varepsilon_{ext}})^\mu], \qquad (3)$$

where $\varepsilon_a$ and $\mu$ are activation field and dynamical exponent. When $\varepsilon_{ext} \leq 1$ MV/cm, the velocity of

PDW$_{(010)}$ are on good agreement with $\mu = 1$, referred to as Merz's law for 2D ferroelectric domain walls[45, 46]. These findings indicate that the experimentally observed coercive field in κ-Ga$_2$O$_3$, its experimentally observed coercive field can be well described by the KAI model[19, 20, 41].

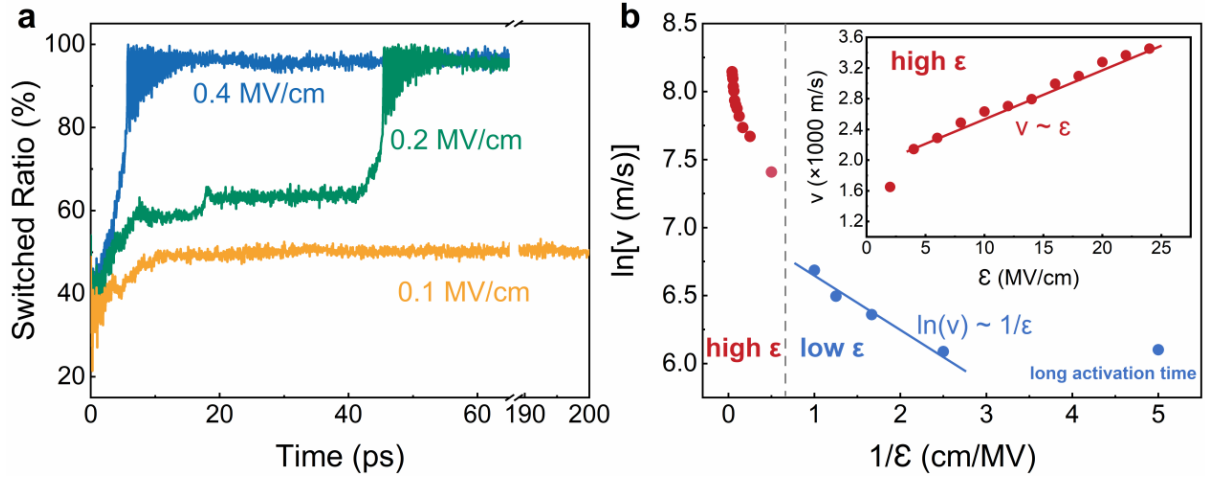

**Figure 4**. **Polarization switching and PDW propagation velocity under different external fields.** (a) Time evolution of the switched domain ratio under different applied electric fields, starting from a pre-nucleated configuration (see Supplementary Note 4 for details). (b) Relationship between electric field and PDW$_{(010)}$ velocity. The insert shows the velocity is proportional to electric field at high-field region (red scatters). When $\varepsilon_{ext} \geq 4$ MV/cm (light blue scatters), the PDW$_{(010)}$ propagation conform to ln(v) ~ $1/\varepsilon_{ext}$, namely Merz's law.

## Influence of 120° lattice domains on polarization switching.

In fact, the single-crystal picture could be a naïve approach for the real κ-Ga$_2$O$_3$ sample. The epitaxially synthesized samples are confirmed to naturally exist the polycrystal structure of in-plane lattice domains rotated by 120° by XRD and TEM experiments, not only in κ-Ga$_2$O$_3$ but also in other compounds with space group $Pna2_1$, e.g., Fe$_2$O$_3$, AlFeO$_3$, and GaFeO$_3$[13, 47, 48, 49, 50, 51, 52]. The lattice constant ration a/b ≈ $1/\sqrt{3}$ also supports this rotational lattice domains. The KAI-type wall motion described above reproduces the experimental coercive field only after a reversed nucleation region is present. To solve the question how real κ-Ga$_2$O$_3$ crystal bypasses the nucleation step seen in our single crystal simulations, therefore, the polycrystal structure is constructed (Fig. 5). The three 120° lattice domains can be labelled as A, B, C (Fig. 5a). Each lattice domain supports two polar states ($\pm P$), yielding six possible ferroelectric variants: $A^+$, $A^-$, $B^+$, $B^-$, $C^+$, $C^-$. We would note that unlike the $Z_3 \times Z_2$ manifold familiar in perovskites[53, 54], here the set {A, B, C} does not arise from a single high-symmetry parent phase. Given that the polarization switching in κ-$Ga_2O_3$ is driven by the in-plane sliding of Ga-O sublayers, at the 120° lattice-domain walls (LDWs) the sliding vector can be discontinuous, or even fully terminated. Therefore, the polarization reversal dynamics can have significant differences than that in the single-crystal picture.

We built a supercell that contains the {A, B, C} lattice domains for DPLR calculations (Fig. 5b). It is a rhombohedral cell with 600 unit cells (lattice constant ~ 17.6 nm). The total atom number is 24,000. We would note that in principle there should be more than one way to construct such a multi-domain structure; the structure we choose here reproduces best the experimental XRD patterns[47, 49], including not only the six-fold, strong (200) diffraction peaks, but also all satellite superlattice peaks around (Fig. 5c). As seen in Fig. 5b, in addition to the A|B|C LDWs, the supercell model naturally hosts two special junctions, junction 1 where three domains meet and junction 2 where six meet. Fig. 5d,5e (see also Supplementary Fig. 9) illustrate the local atomic structure at these junctions together with the sliding vector as the order parameter (cyan arrows) in every adjoining domain. Using the DPLR potential, we compute the internal energy as a function of domain size (Fig. 5f), defined as the long diagonal of the rhombic domain (as marked in Fig. 5b). It reveals that the formation energy of multi-domain structure is relatively low when the domain size exceeds 4 nm. Specifically, for the 18-nm-domain-size model used here, the energy is only 7 meV/f.u. higher than an ideal single-domain crystal, indicating that the simulated structure is thermodynamically stable. The stability is further confirmed by a DP MD run at 500 K (Supplementary Fig. 10).

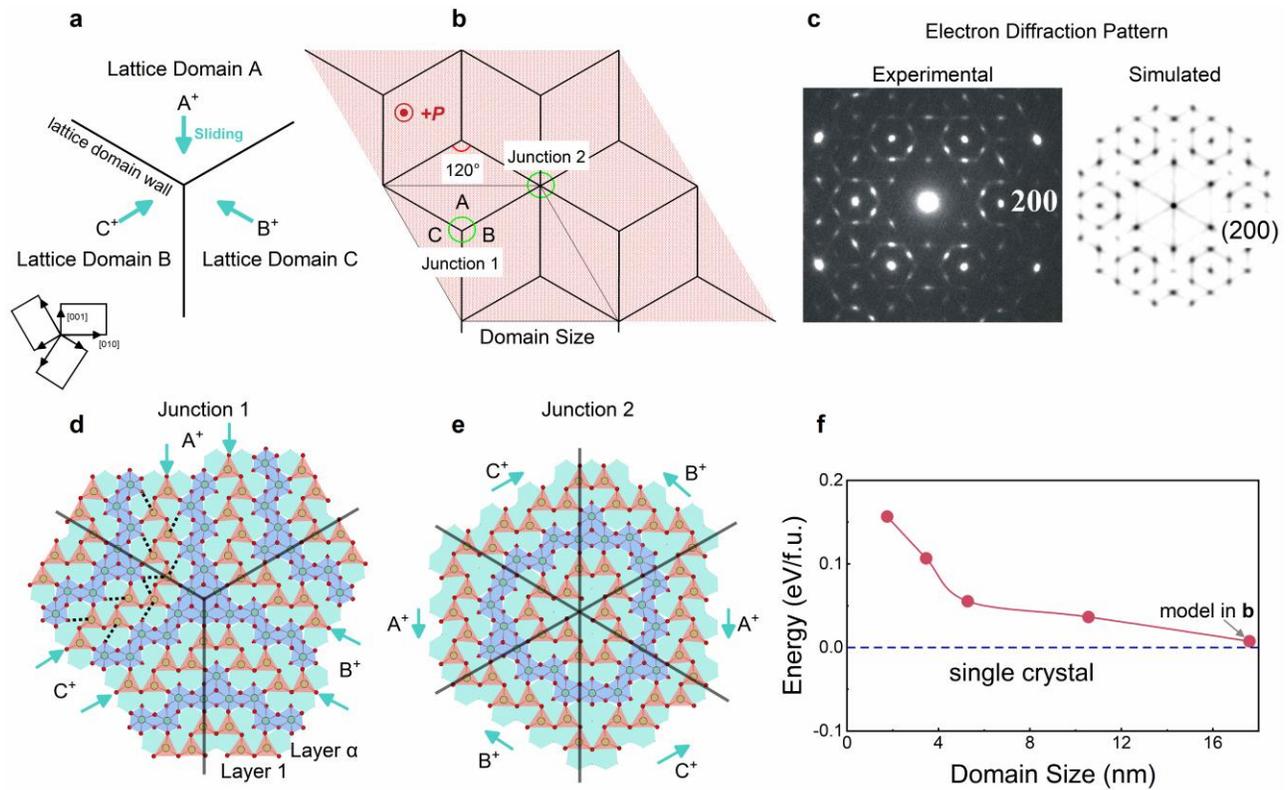

**Figure 5. Atomistic configuration and energy of the 120° rotated lattice domain model in κ-Ga$_2$O$_3$.** (a) Schematic view of the three in-plane lattice domains A, B, C and their 120° LDWs (black lines). Cyan arrows denote the order parameter (sliding vector) in each domain. (b) Large-scale supercell (~ 18 nm) used in DPLR simulations. Black lines are the LDWs and green circles highlight the junctions. (c) The experimental XRD pattern from previous study[49] and the simulated XRD pattern calculated from the supercell in (b), both in [001] projection. The atomic configurations in layer 1 and layer α are depicted near (d) junction 1 and (e) junction 2 (see also Supplementary Fig. 9 for layers 2 and β). (f) Total energy as a function of domain size, where the energy zero is the total energy in an ideal single crystal.

Importantly, it can be proved that PDW propagations cannot cross certain 120° LDW. Fig. 6 illustrates the schematic view. Take the LDW between the A and B variants with the same polar state, $A^+ \parallel B^+$, for which the sliding vectors enclose a 120° angle (Fig. 6a). Suppose an $A^-/A^+$ PDW approaches this LDW from the left, giving the initial sequence $A^-/A^+ \parallel B^+$, where "/" marks the PDW and "∥" the LDW. If the PDW are able to pass through the LDW, the final arrangement would become $A^- \parallel B^-/B^+$ (Fig. 6b). Comparing the two states shows that the LDW itself must switch from $A^+ \parallel B^+$ to $A^- \parallel B^-$. For the A side, changing from $A^+$ to $A^-$ requires the sliding layers to translate by $\sqrt{3}a/6$ perpendicular to the LDW; the same shift is needed for $B^+ \to B^-$ on the right. Let $L_{A^+\|B^+}$ and $L_{A^-\|B^-}$ be the physical widths of LDW before and after switching. Continuity demands

$$L_{A^-\|B^-} = L_{A^+\|B^+} \pm \sqrt{3}a/3, \qquad (4)$$

thus the lattice constant normal to the wall would have to change by $\Delta = \sqrt{3}a/3$. In the limitation where the LDW is atomically sharp ($L_{A^+\|B^+} \to 0$), the required change is equivalent to inserting a finite slice of crystal out of nothing, which is topologically forbidden in a perfect lattice. The same argument holds for LDWs of $B^+ \| C^+$ ($B^- \| C^-$) and $A^+ \| C^+$ ($A^- \| C^-$). These LDWs then act as impenetrable barriers that pin residual PDWs from propagation or immigration.

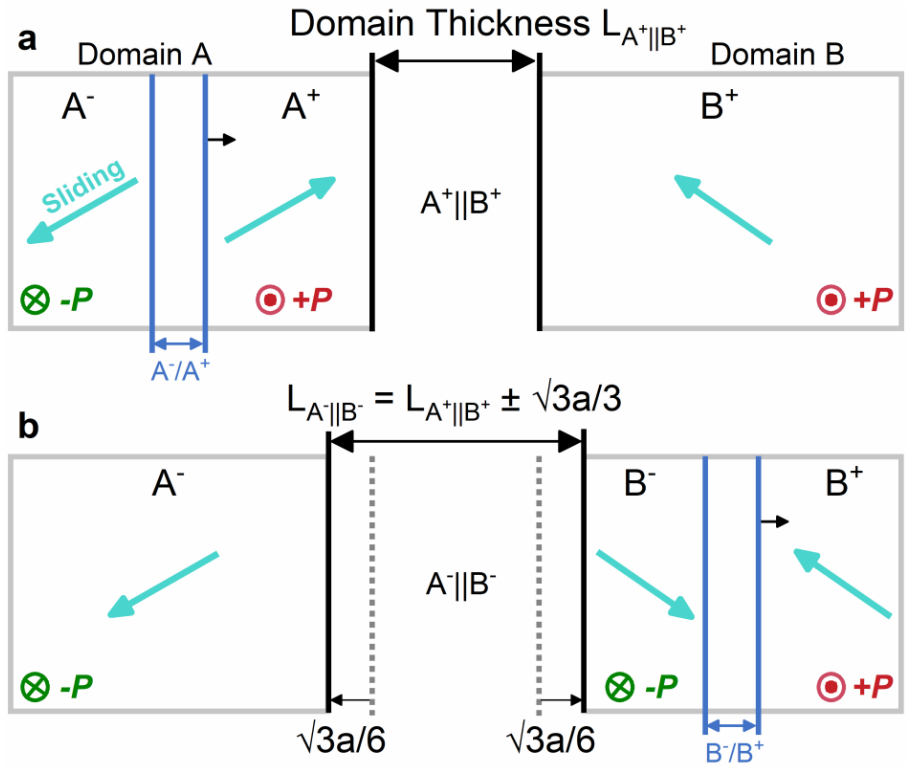

**Figure 6. Schematic diagram of LDW pinning PDW propagation.** (a) Domain configurations with polarization domain $A^-/A^+$ and lattice domain $A^+ \| B^+$. The light blue and black lines are PDW and LDW, respectively. (b) Domain configurations with polarization domain $B^-/B^+$ and lattice domain $A^- \| B^-$. The PDW moves from Domain A ($A^-/A^+$) to Domain B ($B^-/B^+$).

To verify the interaction between PDW and LDW, we simulate the field-driven FE switching in the multidomain supercell with the trained DPLR potential (Fig. 7). Nucleation appears in each lattice domain but, owing to thermal noise, the three nuclei grow at slightly different rates (Fig. 7a). As the domains expand, PDW$_{(100)}$ in three lattice domains approach and touch at ~ 900 fs (Fig. 7b). Before contact, the PDW$_{(100)}$ in domain B moves at 7004 m/s, whereas its perpendicular PDW$_{(010)}$ moves at 3686 m/s, reproducing the single-crystal anisotropy. Along the [100] direction, once a PDW propagates and meets an LDW, its motion stops, exactly as predicted by our argument above. Along the [010] direction, the PDW continues until it reaches junction 2, then halts at a critical distance from junction 2.

After the driving field is removed, the system rapidly settles into a thermodynamically stable network of residual PDWs pinned by LDWs (Fig. 7c,d). The FE-reversal region saturates at 60% of the whole lattice (Fig. 7e), giving a remanent polarization $P = 0.6P_0 = 14.7\ \mu C/cm^2$. We note that it already approaches the experimental values (< 8.6 μC/cm$^2$)[10, 11]. Of course, many factors, such as defects, doping, dislocations, and interfaces, also restrict the remanent polarization, and will cause theoretically predicted values to decrease further. However, these are beyond the scope of the current paper. We would like to emphasize that our results indicates that the size of reversal region, and hence the remanent polarization, can vary with the lattice domain size. By assuming that the non-switchable area around junction 2 is consistent in different domain sizes, the remanent polarization can be approximated as

$$P = \frac{(N_{total} - N_{ns})}{N_{total}} P_0, \tag{5}$$

where $P_0$ is the polarization in primitive cell model, $N_{total}$ is the total number of Ga-O polyhedra in one lattice domain, and $N_{ns}$ is the number of Ga-O polyhedra per lattice domain which are non-switchable around junction 2. We plot in Fig. 7f the approximated remanent polarization for experimental rotational domain size ranging from 5 to 20 nm[6, 47]. For instance, a domain size of 12 nm yields a remanent polarization of ~ 5 μC/cm$^2$, fitting well with experimental values. We therefore suggest that such polarization dependence on domain size may explain the large difference in remanent polarization observed in experiments, with additional reductions expected from defects, dopants, dislocations and interfaces.

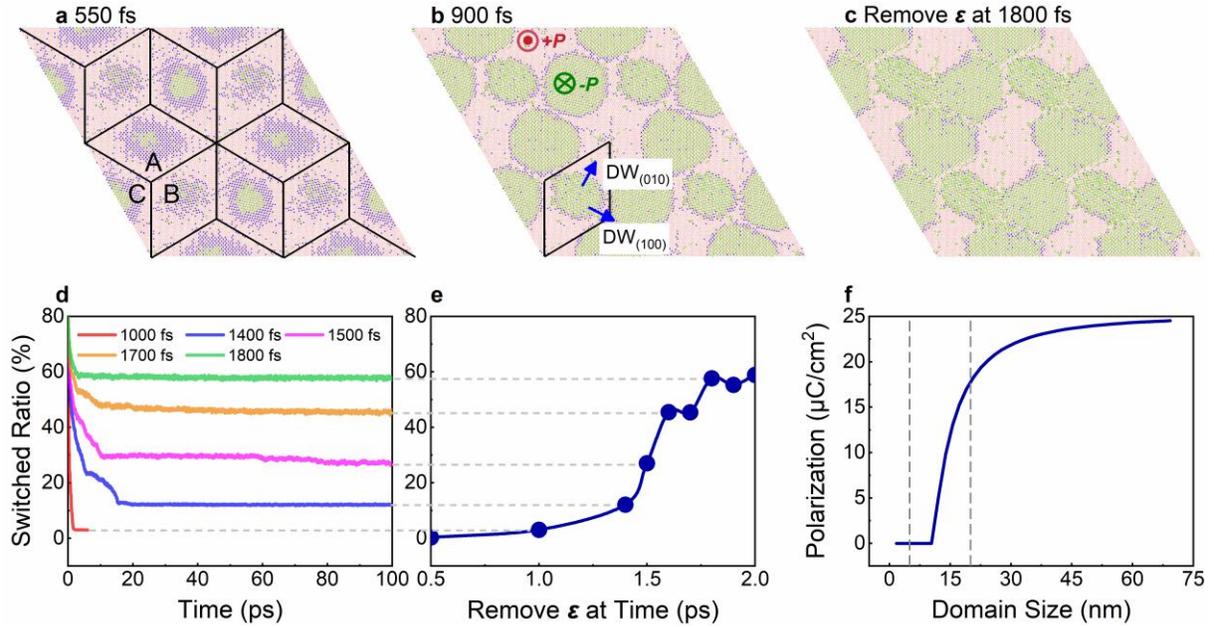

**Figure 7. Field-driven switching and remanent polarization in the $A^+ \parallel B^+ \parallel C^+$ multidomain model.** (a,b) Snapshots of the growing nuclei and PDWs at 550 fs and 900 fs under an electric field along $[00\bar{1}]$. (c) Stable DW pattern after the field is turned off at 1.8 ps; residual PDWs remain pinned at the LDWs. (d) Switched-volume percentage versus time after field removal. Colors denote different duration time of driving field before removal. The system reaches equilibrium within 80 ps. (e) The final switched percentage as a function of the duration time of driving field. (f) Calculated remanent polarization as a function of lattice-domain size; the two vertical lines show the experimental range for domain sizes of 5–20 nm.

During the experimental synthesis process, samples often contain both lattice and polar domains. We therefore construct a supercell with $A^+ \parallel B^- \parallel C^+$ configuration that combines low-barrier PDW motion with LDW pinning (Fig. 8). Its energy is higher than that of either a single crystal or the $A^+ \parallel B^+ \parallel C^+$ configuration, yet AIMD at 500 K shows no structural decay, indicating experimental accessibility (see also Supplementary Fig. 11). The pre-existing PDWs at the $A^+ \parallel B^-$ and $B^- \parallel C^+$ boundaries act as nuclei. Under a 3 MV/cm field, they sweep rapidly across domains A and C, but halt at the $A^+ \parallel C^+$ LDW (Fig. 8a,b). Both the forward switching ($+P \rightarrow -P$) and reverse switching ($-P \rightarrow +P$) finish within 2.5 ps (Fig. 8c). The propagation velocity of PDW$_{(010)}$ is 2056 m/s (see also Supplementary

Fig. 12 for the critical configurations). The larger electric field required for PDW propagation here than that in the single crystal case (~0.4 MV/cm) indicates the pinning effect of LDW. The calculated remanent polarization is 2.61 µC/cm² and rises with lattice domain size.

Because nucleation is supplied by the PDWs which are pinned to LDW, both forward and reverse switching are now governed by KAI-type wall motion (Fig. 8d). We would note that fast wall motions are a hallmark of two-dimensional sliding ferroelectricity, such as 6000 m/s in bilayer h-BN[28] and 3000 m/s in 3R-MoS$_2$[55] with low electric field, while it drops to 300 µm/s in the small-angle twisted WSe$_2$ bilayers[56]. However, the existence of a thermodynamically stable window for sliding-like pathway in 3D bulk compounds remains an open question—one that carries obvious significance for non-volatile ferroelectric memory based on shear-sliding mechanisms.

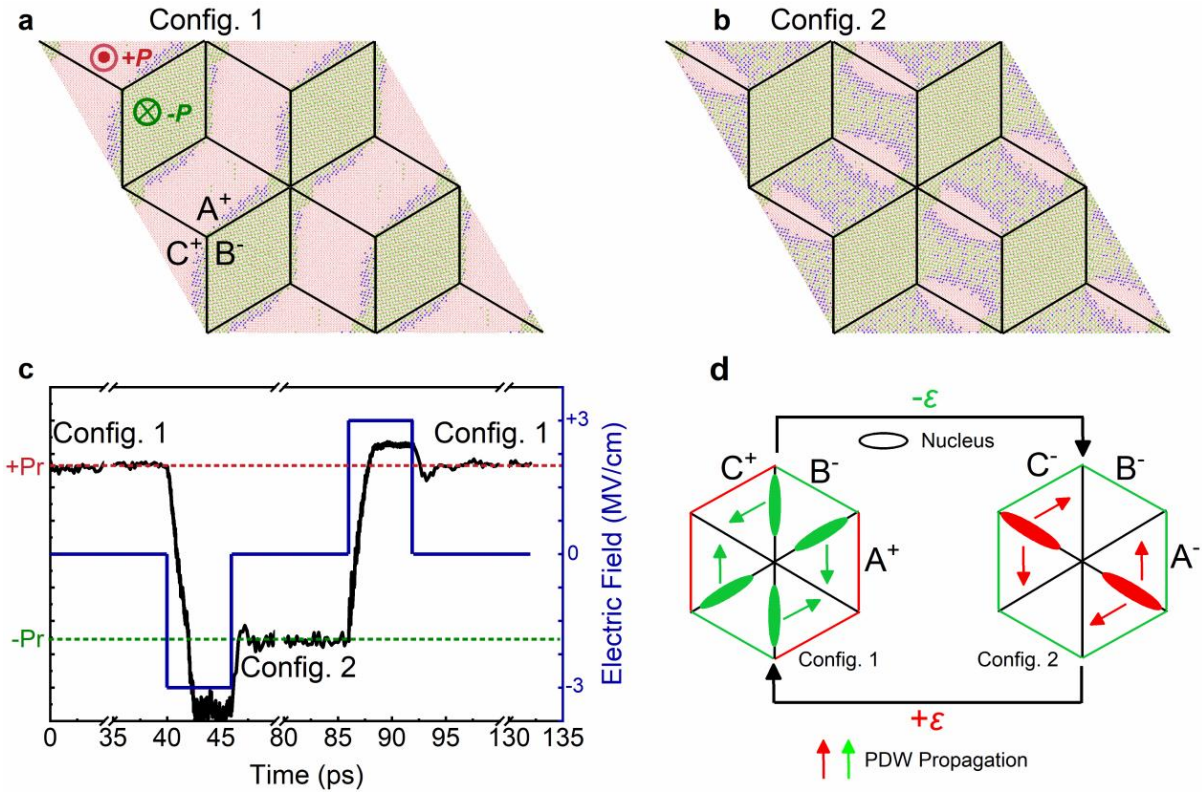

**Figure 8. Field-driven switching in the $A^+ \parallel B^- \parallel C^+$ multidomain model.** (a) Initial domain arrangement with positive remanent polarization (config. 1). (b) Configuration after the driving field (config. 2). (c) Time evolution of remanent polarization under driving electric pulses (max intensity: 3 MV/cm). (d) Schematic of forward ($+P \to -P$) and reverse ($-P \to +P$) switching. Red and green areas indicate +P and −P regions, respectively.

## Discussion

We report a nontrivial ferroelectric switching mechanism from the interplay between polarization domain walls (PDW) and lattice domain walls (LDW), and how it closes the gap between experimental observations and theoretical predictions in the wide-band-gap κ-$Ga_2O_3$. By combining density-functional theory, a long-range machine-learning force field, and large-scale molecular dynamics, we show that the out-of-plane polarization flips through an in-plane shear and sliding of Ga–O layers. Our simulations capture both stages of field-driven switching: slow nucleation followed by rapid domain-wall motion. Two types of PDW appear. The (100) wall travels twice as fast as the (010) wall, the latter of which has a velocity following Merz's law at low fields. These data reconcile the experimentally low coercive field with the Kolmogorov–Avrami–Ishibashi model. The three 120° LDWs that exist in real samples can topologically terminate PDW propagation. This pinning effect leaves a network of residual domain walls, supplies spontaneous nucleation region, and hence enhances rapid FE switching under low field while suppresses the remanent polarization to a fraction of its intrinsic value. Our results enrich the domain engineering in this family of FE materials for fast, low-power ferroelectric devices, and provide a transferable framework for the study on other ferroelectric systems.

## Methods

### DFT and MD Simulations

The DFT simulations are performed by Vienna ab initio Simulation Package (VASP)[58], Quickstep/CP2K package[59], and PWmat[60, 61]. In all calculations, the revised Perdew-Burke-Erzerhof (PBEsol) exchange-correlation functional is adopted.[62] Ga($3d^{10}4s^24p^1$) and O($2s^22p^4$) are treated as valence electrons. In the Quickstep/CP2K simulations, the DZVP-MOLOPT-SR-GTH basis sets and Goedecker-Teter-Hutter (GHT) potentials[63, 64] is employed to get Molecular Dynamics (MD) trajectories. A 2×2×2 supercell, containing 320 atoms, is constructed from the unitcell. The canonical ensemble is used in all MD simulations with 1 fs time step at the Γ point. The BECs and MD with finite periodic electric field compute using the Berry phase approach[37, 38, 64]. The Wannier centers are calculated from the maximally localized Wannier functions[65]. Configurations are randomly chosen and incorporated into the machine learning database. In order to improve the accuracy of the deep potential model, VASP is used to obtain energy and atomic force of configurations in database. Valence electron-ion core interactions are treated by projected-augmented wave method[66]. The plane wave energy cutoff of 550 eV and a 2×1×1 Γ-centered k-point mesh is applied. The energy convergence criterion is set to 0.001 meV. The energy barrier is calculated by climbing image nudged elastic band (CI-NEB) methods[14], implemented within the PWmat code. The calculations are based on norm-conserving pseudopotentials[67]. The classical MD simulations are carried out by LAMMPS code[68], which is modified to support deep potential.

**Machine Learning Force Field**

The standard deep potential (DP) is to represent the total energy of the systems as a sum of atomic contributions determined by the local environment with a cutoff radius, $E = \sum_i E_i$[35, 69]. To incorporate the explicit long-rang interactions into the model, the method of deep potential long-range (DPLR) is employed to obtain the force field[22]. This model approximates the long-range electrostatic interaction between ions (nuclei + core electrons) and valence electrons with that of distributions of spherical Gaussian charges located at ionic and electronic sites.[32] In this model, total energy is derived from two parts ($E = E_{sr} + E_{lr}$): one part is short-long interactions calculated from standard DP network and long-rang electrostatic interactions with Ewald method calculated form deep Wannier (DW) network[36]. The DPLR embedded in DeePMD-kit code[32, 33] is trained in two steps: initially, the DW model is trained, and the the DPLR model is trained. In high ionic materials, the Wannier centers (WCs) are located around the atomic nuclei. Different types of atoms have different numbers of WCs. For example, there are 5 WCs around Ga ions and 4 WCs around O ions. Based on this fact, the average WC only related to ionic position ($R_i$) is defined as $\omega_i = 1/n \sum_j \omega_j - R_i$. The particle-particle-particle-mesh (PPPM) method for calculating the electrostatic energy is used in this algorithm. Importantly, the WCs also can calculate the electronic part of polarization of system, $P = P_{ion} + P_{ele}$. With an electric field ($\varepsilon_{ext}$), the total energy can be calculated by $E = E_{DPLR} - P \cdot \varepsilon_{ext}$, which is similar to electric enthalpy functional[38, 70]. To train these models, a descriptor of DeePot-SE (including radial and angular information of atomic configuration) is adopted[32], and a three-layer embedding net (25, 50, 100) and a three-layer fitting net (240, 240, 240) are chose. The descriptor characterizes the local environment of an atom within a cutoff radius set to 8.5 Å. A more detailed description of the process and algorithm can be found on the homepage of DeePMD-kit and corresponding references.

**Notes**

The authors declare no competing financial interest.

**ACKNOWLEDGEMENTS**

This work is supported by the National Key R&D Program of China (2022YFB3605400). Y.Z. is supported by the Postdoctoral Fellowship Program of CPSF (GZB20240720) and Project funded by China Postdoctoral Science Foundation (2024M763182). Z.W. is supported by the National Natural Science Foundation of China (12174380).

# REFERENCES

1. He L, Özdemir ŞK, Zhu J, Kim W, Yang L. Detecting single viruses and nanoparticles using whispering gallery microlasers. *Nat. Nanotechnol.* **6**, 428-432 (2011).

2. Cui B, *et al.* Ferroelectric photosensor network: an advanced hardware solution to real-time machine vision. *Nat. Commun.* **13**, 1707 (2022).

3. Fang M, *et al.* Tuning the interfacial spin-orbit coupling with ferroelectricity. *Nat. Commun.* **11**, 2627 (2020).

4. Scott JF, Paz de Araujo CA. Ferroelectric memories. *Science* **246**, 1400-1405 (1989).

5. Kneiß M, *et al.* Tin-assisted heteroepitaxial PLD-growth of κ-$Ga_2O_3$ thin films with high crystalline quality. *APL Mater.* **7**, 022516 (2018).

6. Mazzolini P, *et al.* Silane-mediated expansion of domains in Si-doped κ-$Ga_2O_3$ epitaxy and its impact on the in-plane electronic conduction. *Adv. Funct. Mater.* **33**, 2207821 (2023).

7. Cho SB, Mishra R. Epitaxial engineering of polar ε-$Ga_2O_3$ for tunable two-dimensional electron gas at the heterointerface. *Appl. Phys. Lett.* **112**, 162101 (2018).

8. Maccioni MB, Fiorentini V. Phase diagram and polarization of stable phases of $(Ga_{1-x}In_x)_2O_3$. *Appl. Phys. Express* **9**, 041102 (2016).

9. Shimada K. First-principles study of crystal structure, elastic stiffness constants, piezoelectric constants, and spontaneous polarization of orthorhombic $Pna2_1$-$M_2O_3$ (M = Al, Ga, In, Sc, Y). *Mater. Res. Express* **5**, 036502 (2018).

10. Rao BN, *et al.* Investigation of ferrimagnetism and ferroelectricity in $Al_xFe_{2-x}O_3$ thin films. *J. Mater. Chem. C* **8**, 706-714 (2020).

11. Katayama T, Yasui S, Hamasaki Y, Osakabe T, Itoh M. Chemical tuning of room-temperature ferrimagnetism and ferroelectricity in ε-$Fe_2O_3$-type multiferroic oxide thin films. *J. Mater. Chem. C* **5**, 12597-12601 (2017).

12. Ducharme S, *et al.* Intrinsic ferroelectric coercive field. *Phys. Rev. Lett.* **84**, 175-178 (2000).

13. Janzen BM, *et al.* Comprehensive Raman study of orthorhombic κ/ε-$Ga_2O_3$ and the impact of rotational domains. *J. Mater. Chem. C* **9**, 14175-14189 (2021).

14. Henkelman G, Uberuaga BP, Jónsson H. A climbing image nudged elastic band method for finding saddle points and minimum energy paths. *J. Chem. Phys.* **113**, 9901-9904 (2000).

15. King-Smith RD, Vanderbilt D. Theory of polarization of crystalline solids. *Phys. Rev. B* **47**, 1651-1654 (1993).

16. Tao A, *et al.* Ferroelectric polarization and magnetic structure at domain walls in a multiferroic film. *Nat. Commun.* **15**, 6099 (2024).

17. Xu K, Feng JS, Liu ZP, Xiang HJ. Origin of ferrimagnetism and ferroelectricity in room-temperature multiferroic ε-$Fe_2O_3$. *Phys. Rev. Appl.* **9**, 044011 (2018).

18. Momma K, Izumi F. VESTA3 for three-dimensional visualization of crystal, volumetric and morphology data. *J. Appl. Crystallogr.* **44**, 1272-1276 (2011).

19. Liu S, Grinberg I, Rappe AM. Intrinsic ferroelectric switching from first principles. *Nature* **534**, 360-363 (2016).

20. Shin Y-H, Grinberg I, Chen IW, Rappe AM. Nucleation and growth mechanism of ferroelectric domain-wall motion. *Nature* **449**, 881-884 (2007).

21. Unke OT, *et al.* Machine learning force fields. *Chem. Rev.* **121**, 10142-10186 (2021).

22. Zhang L, Wang H, Muniz MC, Panagiotopoulos AZ, Car R, E W. A deep potential model with long-range electrostatic interactions. *J. Chem. Phys.* **156**, 124107 (2022).

23. Staacke CG, Heenen HH, Scheurer C, Csányi G, Reuter K, Margraf JT. On the role of long-range electrostatics in machine-learned interatomic potentials for complex battery materials. *ACS Appl. Energy Mater.* **4**, 12562-12569 (2021).

24. Deng Y, Fu S, Guo J, Xu X, Li H. Anisotropic collective variables with machine learning potential for ab initio crystallization of complex ceramics. *ACS Nano* **17**, 14099-14113 (2023).

25. Gigli L, Veit M, Kotiuga M, Pizzi G, Marzari N, Ceriotti M. Thermodynamics and dielectric response of $BaTiO_3$ by data-driven modeling. *npj Comput. Mater.* **8**, 209 (2022).

26. Myung CW, Hajibabaei A, Cha J-H, Ha M, Kim J, Kim KS. Challenges, opportunities, and prospects in metal halide perovskites from theoretical and machine learning perspectives. *Adv. Energy Mater.* **12**, 2202279 (2022).

27. Zhong W, Vanderbilt D, Rabe KM. First-principles theory of ferroelectric phase transitions for perovskites: The case of $BaTiO_3$. *Phys. Rev. B* **52**, 6301-6312 (1995).

28. He R, *et al.* Ultrafast switching dynamics of the ferroelectric order in stacking-engineered ferroelectrics. *Acta Mater.* **262**, 119416 (2024).

29. Wu J, Zhang Y, Zhang L, Liu S. Deep learning of accurate force field of ferroelectric $HfO_2$. *Phys. Rev. B* **103**, 024108 (2021).

30. He R, Wang H, Liu F, Liu S, Liu H, Zhong Z. Unconventional ferroelectric domain switching dynamics in $CuInP_2S_6$ from first principles. *Phys. Rev. B* **108**, 024305 (2023).

31. Gong Z, Liu JZ, Ding X, Sun J, Deng J. Strain-aided room-temperature second-order ferroelectric phase transition in monolayer PbTe: Deep potential molecular dynamics simulations. *Phys. Rev. B* **108**, 134112 (2023).

32. Wang H, Zhang L, Han J, E W. DeePMD-kit: A deep learning package for many-body potential energy representation and molecular dynamics. *Comput. Phys. Commun.* **228**, 178-184 (2018).

33. Zeng J, *et al.* DeePMD-kit v2: A software package for deep potential models. *J. Chem. Phys.* **159**, 054801 (2023).

34. Gao A, Remsing RC. Self-consistent determination of long-range electrostatics in neural network potentials. *Nat. Commun.* **13**, 1572 (2022).

35. Behler J, Parrinello M. Generalized neural-network representation of high-dimensional potential-energy surfaces. *Phys. Rev. Lett.* **98**, 146401 (2007).

36. Zhang L, Chen M, Wu X, Wang H, E W, Car R. Deep neural network for the dielectric response of insulators. *Phys. Rev. B* **102**, 041121 (2020).

37. Umari P, Pasquarello A. Ab initio molecular dynamics in a finite homogeneous electric field. *Phys. Rev. Lett.* **89**, 157602 (2002).

38. Souza I, Íñiguez J, Vanderbilt D. First-principles approach to insulators in finite electric fields. *Phys. Rev. Lett.* **89**, 117602 (2002).

39. Spaldin NA. A beginner's guide to the modern theory of polarization. *J. Solid State Chem.* **195**, 2-10 (2012).

40. Tagantsev AK, Stolichnov I, Setter N, Cross JS, Tsukada M. Non-Kolmogorov-Avrami switching kinetics in ferroelectric thin films. *Phys. Rev. B* **66**, 214109 (2002).

41. Ishibashi Y, Takagi Y. Note on ferroelectric domain switching. *J. Phys. Soc. Jpn.* **31**, 506-510 (1971).

42. Yang J, *et al.* Theoretical lower limit of coercive field in ferroelectric hafnia. *Phys. Rev. X* **15**, 021042 (2025).

43. Jo JY, *et al.* Nonlinear dynamics of domain-wall propagation in epitaxial ferroelectric thin films. *Phys. Rev. Lett.* **102**, 045701 (2009).

44. Tybell T, Paruch P, Giamarchi T, Triscone JM. Domain wall creep in epitaxial ferroelectric Pb(Zr$_{0.2}$Ti$_{0.8}$)O$_3$ thin films. *Phys. Rev. Lett.* **89**, 097601 (2002).

45. Merz WJ. Domain formation and domain wall motions in ferroelectric BaTiO$_3$ single crystals. *Phys. Rev.* **95**, 690-698 (1954).

46. Miller RC, Weinreich G. Mechanism for the sidewise motion of 180° domain walls in barium titanate. *Phys. Rev.* **117**, 1460-1466 (1960).

47. Cora I, et al. The real structure of ε-Ga$_2$O$_3$ and its relation to κ-phase. *CrystEngComm* **19**, 1509-1516 (2017).

48. Hrubišák F, et al. Heteroepitaxial growth of Ga$_2$O$_3$ on 4H-SiC by liquid-injection MOCVD for improved thermal management of Ga$_2$O$_3$ power devices. *J. Vac. Sci. Technol. A* **41**, 042708 (2023).

49. Cora I, Fogarassy Z, Fornari R, Bosi M, Rečnik A, Pécz B. In situ TEM study of κ→β and κ→γ phase transformations in Ga$_2$O$_3$. *Acta Mater.* **183**, 216-227 (2020).

50. Nishinaka H, Ueda O, Ito Y, Ikenaga N, Hasuike N, Yoshimoto M. Plan-view TEM observation of a single-domain κ-Ga$_2$O$_3$ thin film grown on ε-GaFeO$_3$ substrate using GaCl$_3$ precursor by mist chemical vapor deposition. *Jpn. J. Appl. Phys.* **61**, 018002 (2022).

51. Katayama T, Yasui S, Hamasaki Y, Itoh M. Control of crystal-domain orientation in multiferroic Ga$_{0.6}$Fe$_{1.4}$O$_3$ epitaxial thin films. *Appl. Phys. Lett.* **110**, 212905 (2017).

52. Hamasaki Y, Yasui S, Katayama T, Kiguchi T, Sawai S, Itoh M. Ferroelectric and magnetic properties in ε-Fe$_2$O$_3$ epitaxial film. *Appl. Phys. Lett.* **119**, 182904 (2021).

53. Huang FT, et al. Domain topology and domain switching kinetics in a hybrid improper ferroelectric. *Nat. Commun.* **7**, 11602 (2016).

54. Han M-G, et al. Ferroelectric switching dynamics of topological vortex domains in a hexagonal manganite. *Adv. Mater.* **25**, 2415-2421 (2013).

55. Bian R, et al. Developing fatigue-resistant ferroelectrics using interlayer sliding switching. *Science* **385**, 57-62 (2024).

56. Ko K, et al. Operando electron microscopy investigation of polar domain dynamics in twisted van der Waals homobilayers. *Nat. Mater.* **22**, 992-998 (2023).

57. Buragohain P, Erickson A, Mimura T, Shimizu T, Funakubo H, Gruverman A. Effect of film microstructure on domain nucleation and Intrinsic switching in ferroelectric Y:HfO$_2$ thin film capacitors. *Adv. Funct. Mater.* **32**,

2108876 (2022).

58. Kresse G, Furthmüller J. Efficient iterative schemes for ab initio total-energy calculations using a plane-wave basis set. *Phys. Rev. B* **54**, 11169-11186 (1996).

59. Kühne TD, *et al.* CP2K: An electronic structure and molecular dynamics software package-Quickstep: Efficient and accurate electronic structure calculations. *J. Chem. Phys.* **152**, 194103 (2020).

60. Jia W, *et al.* Fast plane wave density functional theory molecular dynamics calculations on multi-GPU machines. *J. Comput. Phys.* **251**, 102-115 (2013).

61. Jia W, *et al.* The analysis of a plane wave pseudopotential density functional theory code on a GPU machine. *Comput. Phys. Commun.* **184**, 9-18 (2013).

62. Perdew JP, *et al.* Restoring the density-gradient expansion for exchange in solids and surfaces. *Phys. Rev. Lett.* **100**, 136406 (2008).

63. Goedecker S, Teter M, Hutter J. Separable dual-space Gaussian pseudopotentials. *Phys. Rev. B* **54**, 1703-1710 (1996).

64. VandeVondele J, Hutter J. Gaussian basis sets for accurate calculations on molecular systems in gas and condensed phases. *J. Chem. Phys.* **127**, 114105 (2007).

65. Marzari N, Mostofi AA, Yates JR, Souza I, Vanderbilt D. Maximally localized Wannier functions: Theory and applications. *Rev. Mod. Phys.* **84**, 1419-1475 (2012).

66. Blöchl PE. Projector augmented-wave method. *Phys. Rev. B* **50**, 17953-17979 (1994).

67. Hamann DR. Optimized norm-conserving Vanderbilt pseudopotentials. *Phys. Rev. B* **88**, 085117 (2013).

68. Thompson AP, *et al.* LAMMPS - a flexible simulation tool for particle-based materials modeling at the atomic, meso, and continuum scales. *Comput. Phys. Commun.* **271**, 108171 (2022).

69. Zhang L, Han J, Wang H, Car R, E W. Deep potential molecular dynamics: A scalable model with the accuracy of quantum mechanics. *Phys. Rev. Lett.* **120**, 143001 (2018).

70. Nunes RW, Gonze X. Berry-phase treatment of the homogeneous electric field perturbation in insulators. *Phys. Rev. B* **63**, 155107 (2001).